\documentclass[prl,superscriptaddress,showpacs,twocolumn]{revtex4}
\usepackage{graphicx}
\def\figwidth{8cm}
\def\iomn{i\omega_n}
\def\ep{\epsilon_{\perp}}
\newcommand{\cDMFT}{ch-DMFT\,}
\def\tp{t_{\perp}}
\def\kp{k_{\perp}}
\def\kro{K_{\rho}}
\newcommand{\mr}[1]{{{\mathrm{#1}}}}
\def\sp{\sigma_{\perp}}
\begin{document}
\title{Deconfinement transition and Luttinger to Fermi Liquid crossover 
in quasi one-dimensional systems}
\author{S. Biermann}
\affiliation{Laboratoire de Physique
des Solides, CNRS-UMR 8502, UPS B\^at. 510, 91405 Orsay France}
\affiliation{LPTENS CNRS UMR 8549 24, Rue
Lhomond 75231 Paris Cedex 05, France}
\author{A. Georges}
\affiliation{LPTENS CNRS UMR 8549 24, Rue
Lhomond 75231 Paris Cedex 05, France}
\affiliation{Laboratoire de Physique
des Solides, CNRS-UMR 8502, UPS B\^at. 510, 91405 Orsay France}
\author{A. Lichtenstein}
\affiliation{University of Nijmegen, NL-6525 ED Nijmegen, The Netherlands}
\author{T. Giamarchi}
\affiliation{Laboratoire de Physique
des Solides, CNRS-UMR 8502, UPS B\^at. 510, 91405 Orsay France}
\affiliation{LPTENS CNRS UMR 8549 24, Rue
Lhomond 75231 Paris Cedex 05, France}
\date{\today}
\begin{abstract}
We investigate a system of one dimensional Hubbard chains of interacting
fermions coupled by inter-chain hopping. Using a generalization of the Dynamical Mean Field Theory
we study the deconfinement transition from a Mott insulator to a metal
and the crossover between 
Luttinger and Fermi liquid phases. One-particle properties, local spin 
response and inter-chain optical conductivity are calculated. 
Possible applications to organic conductors are discussed.
\end{abstract}
\maketitle

The nature of the metallic phase of interacting electron systems
depends strongly on dimensionality.
In three dimensions, Fermi liquid (FL)
theory applies, whereas
in one dimension the quasi-particle concept breaks down, leading to
a different kind of low-energy fixed point known as a Luttinger
liquid (LL). For commensurate electron fillings, strong
enough repulsive interactions
destroy the metallic state altogether by
opening a Mott gap. This phenomenon exists in all dimensions but
the one-dimensional case is particularly favorable \cite{voit_bosonization_revue}.
In quasi one-dimensional (Q1D) systems, interchain hopping 
can induce a (deconfinement) transition from the Mott insulating (MI)
state to a metallic state and crossovers between
different metallic behaviors (Fig.~1).
Besides their intrinsic theoretical interest, understanding 
these phenomena
is directly relevant for a number of compounds
such as the organic (super)-conductors (Bechgaard salts),
which are three dimensional stacks of quarter- filled
chains\cite{jerome_revue_1d}. Indeed, in these compounds some of the low
temperature properties are well described by Fermi liquid
theory, whereas optical \cite{schwartz_electrodynamics} and
transport properties \cite{moser_conductivite_1d}
have shown that the high temperature phase is either
a Luttinger liquid or a Mott insulator.
\begin{figure}
\centerline{\includegraphics[width=\figwidth]{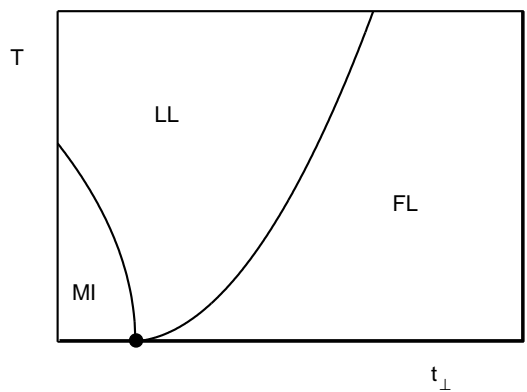}}
\caption{\label{Phase_diagram} Schematic phase diagram of a quasi
one-dimensional system, as a function of temperature $T$ and
inter-chain hopping $\tp$. The Mott-Insulator (MI), Luttinger Liquid
(LL) and Fermi Liquid (FL) regimes are displayed. All lines are
crossovers, except for the $T=0$ transition from a MI to a FL. The
MI phase is present only for commensurate fillings. Possible
phases with long-range order have been omitted.}
\end{figure}

Describing these Q1D systems is not an easy task. The transverse
hopping $t_\perp$ is a relevant perturbation on the LL
\cite{bourbonnais_rmn,wen_coupled_chains,yakovenko_manychains}.
Hence, perturbative renormalization group calculations yield an 
estimate of the crossover scale \cite{bourbonnais_rmn,bourbonnais_couplage}
but fail below that scale. In the MI state,
electrons are confined on the chains by the Mott gap. A finite
critical value of $t_\perp$ is needed to induce an insulator
to metal transition. Such a transition has been advocated
\cite{giamarchi_mott_shortrev,vescoli_confinement_science}to be at the root of
the change from insulating to metallic behavior observed in
Bechgaard salts when increasing pressure or when going from the
TMTTF to the TMTSF family.
Thus the effects of the inter-chain hopping that are the most
important physically 
{\it cannot be handled reliably by perturbative methods}.
Although some non-perturbative studies
of deconfinement have been made for a finite number of
chains~\cite{tsuchiizu_confinement_spinful_refs}
the case of an infinite system is still open.
Some of the key questions yet to be answered are: (i) What is the
crossover scale from the LL to MI, or from the LL to FL (Fig.1), and is
there only one crossover scale for the different physical
properties (transport, spin response, single-particle
properties etc...) ?
(ii) What is the nature of the
low-temperature FL state, and in particular is the shape of the
Fermi surface (FS) affected by interactions ? Do quasi-particle
(QP) properties such as the quasi-particle residue $Z(k)$, vary
significantly as the wavevector $k$ moves along the FS ?

In order to deal with these questions in a non-perturbative manner,
a new method has been
proposed~\cite{arrigoni_tperp_resummation,georges_organics_dinfiplusone}.
This method, which generalizes the single site dynamical
mean-field theory (DMFT)~\cite{georges_d=infini},
replaces the quasi one-dimensional system by a single {\it
effective chain} from which electrons can hop to a self-consistent
bath. We thus nickname it {\it chain-DMFT} (\cDMFT).
It is exact in
the limit of decoupled chains, while it reduces to usual DMFT in the opposite limit
of zero-hopping along the chains, hence capturing the physics of a
strongly correlated FL (or MI) ~\cite{georges_d=infini}. Solving
the mean field equations is a very challenging task however.
In this paper, the first quantitative solution of the \cDMFT
equations is presented. We show that the above issues can
be valuably addressed, and some of the above questions
answered.
In view of the difficulty of
the problem only the Hubbard model is considered in this paper,
but more realistic extensions are in sight, as discussed below.

Let us consider a system of coupled chains described by a sum of Hamiltonians
of the isolated chains plus a hopping term between neighboring chains.
The \cDMFT approximates this system by the effective action:
\begin{eqnarray}
\nonumber S_{\text{eff}}=&& - \int\int^{\beta}_{0}d\tau\,d\tau'
\sum_{ij,\sigma} c^{+}_{i\sigma}(\tau) {\cal
G}_0^{-1}(i-j,\tau-\tau')
c_{j\sigma}(\tau')\\
&&+\int_0^{\beta} d\tau
H^{int}_{1D}[\{c_{i\sigma},c^+_{i\sigma}\}] \label{Seff}
\end{eqnarray}
where $H^{int}_{1D}$ is the interacting part of the on-chain
Hamiltonian.
${\cal G}_0$ is an {\it
effective propagator} determined from a self-consistency
condition: the Green's function
$G(i-j,\tau-\tau')\equiv -\langle c(i,\tau)c^+(j,\tau')
\rangle_{\text{eff}}$ calculated from $S_{\text{eff}}$ should coincide
with the on-chain Green's function of the original problem, with
the same self-energy $\Sigma={\cal G}_0^{-1}-G^{-1}$. This reads:
\begin{equation}\label{sc_cond}
G(k,\iomn) = \int d\ep
{{D(\ep)}\over{\iomn+\mu-\epsilon_k-\Sigma(k,\iomn)-\ep}}
\end{equation}
Let $\ep(\kp)$
denote the Fourier transform of the
inter-chain hopping $\tp^{mm'}$,
$D(\ep)=\sum_{k\perp}\delta[\ep-\ep(\kp)]$ the corresponding
density of states,
$k$ the momentum in the chain direction and $\omega_n$ the
Matsubara frequencies corresponding to the inverse temperature
$\beta$.
The \cDMFT equations
(\ref{Seff},\ref{sc_cond}) fully determine the self-energy and
Green's function of the coupled chains
$G(k,\kp,\iomn)^{-1}=\iomn+\mu-\epsilon_k-\epsilon_\perp(\kp)-\Sigma(k,\iomn)$.
The \cDMFT approach can be rigorously
justified in the formal limit where the lattice connectivity in
the transverse direction is taken to infinity.
In systems with finite transverse connectivity it can be viewed
as an approximation neglecting the dependence
of the self-energy on transverse momentum, keeping both frequency
and in-chain momentum dependence: $\Sigma=\Sigma(k,\iomn)$.
In the following, we specialize to the Hubbard model, i.e for each
chain:
\begin{equation}
H_{1D}=-t\sum_{i}\,\left[c^+_ic_{i+1}+c^+_{i+1}c_i\right]\,+\,
U\sum_i\,n_{i\uparrow}n_{i\downarrow}
\end{equation}
and we consider a two-dimensional array of chains forming a square
lattice with nearest neighbor hopping, so that the transverse
dispersion is $\epsilon_\perp(\kp)\,=-2\tp\,\cos\kp$.

We solve the effective chain problem for chains of 16 or 32 sites
with periodic boundary conditions using the QMC Hirsch-Fye
algorithm \cite{hirsch_qmc}. Using 32
time-slices in imaginary time allows to reliably access
temperatures down to $T/W\simeq 1/50=0.02$ with $U/W\simeq 1$
where $W=4t$ is the bandwidth of the one-dimensional chain.
Typically, about 5000 QMC sweeps and 10 to 15 \cDMFT iterations
are sufficient to reach convergence.
The main quantities that we calculate and analyze are: a) On-chain
(i.e summed over $\kp$) single particle Green's functions:
$G(k,\iomn)$, and self-energies $\Sigma(k,\iomn)$ (obtained from
${\cal G}_0^{-1}-G^{-1}$
\footnote{To get a reliable estimate for the self-energy
we used an analytical Fourier transform of a spline interpolated
Green's function with exact boundary
conditions in the spirit of
Joo et al., condmat-0009367.}). This allows to  identify
the location of the FS $\kp^F(k)$ by solving
$\epsilon_\perp(\kp^F)=
\mu-{\rm Re}\Sigma(k,i\omega_{n=1})-\epsilon_k$ and the
QP residue $Z=Z(\kp^F)$ by fitting the slope ($=1-Z^{-1}$) of
$\Sigma$ vs. $\iomn$ at a specific FS point.
b) On-chain spin and charge response functions and
in particular the local
spin response function: $\chi_s(\tau)=\langle
S^z(j,0)S^z(j,\tau)\rangle =
\sum_{k,k\perp}\chi_s(k,k_{\perp},\tau)$. Indeed, in a LL (or
with $\kro=1$ in a FL) we have
$\chi_s(\tau)=\chi_s(\beta/2)\,
\left(\sin\pi\tau/\beta \right)^{-(1+\kro)}$
in the asymptotic regime where $\beta,\tau$ are larger than some
coherence scale. By fitting our results to this form, we can define an
effective $\kro$.
Furthermore, the temperature dependence of
$\chi_s(\beta/2)$ is closely related to the NMR relaxation rate
$1/T_1\equiv \lim_{\omega \to 0}\chi_s''(\omega)/\omega$. In a FL
liquid at low-enough temperature \cite{randeria_imaginary_correlations}: $1/T_1 =
\beta\,\chi_s(\beta/2)/{2 \pi^2} $.
In a LL
$1/T_1$ and $\beta\,\chi_s(\beta/2)$
have the same
T-dependence (albeit with a proportionality factor depending on
$\kro$). c) Inter-chain optical conductivity. Within \cDMFT
vertex corrections drop out \cite{georges_organics_dinfiplusone}
and the conductivity is given by
\begin{eqnarray}
\nonumber \sp (i\omega) \propto &&\frac{\tp^2}{\omega}\int
{{d\kp}\over{2\pi}}\, \sin^2\kp\,
\int {{dk}\over{2\pi}}\,\frac{1}{\beta}\sum_m\,G(k,\kp,i\omega_m)\\
&&\times G(k,\kp,i\omega_m+i\omega)\label{sigmatrue}
\end{eqnarray}
Note that we have taken into account the $\kp$- dependence of the
current vertex in (\ref{sigmatrue}). In practice, we perform
analytically the $\kp$-integration in (\ref{sigmatrue}) and
numerically the k-integration and Matsubara summation. We then use
the Maximum Entropy (MaxEnt) algorithm \cite{jarrell_maxent}
to continue $\sp(i\omega)$ to the real axis and obtain the
inter-chain optical conductivity $\mr{Re}\,\sp(\omega,T)$.

\begin{figure}
\centerline{\includegraphics[width=\figwidth]{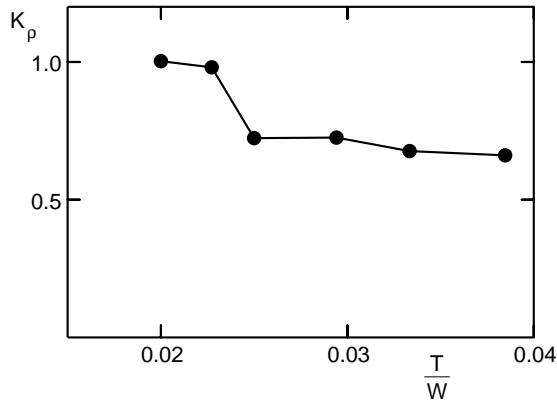}}
\caption{\label{kro-vs-T} Effective $\kro$ {\it vs.} temperature
in the doped case (filling $n \simeq 0.8$) for $U/W=1.0$,
$\tp/W=0.14$.}
\end{figure}

We first discuss our results away from half-filling, for $U/W=1$
and $\mu=0.2$ (corresponding to a total density $n\simeq 0.8$).
The numerical method was thoroughly tested for decoupled
chains ($\tp=0$), for which our fitting procedure of $\chi_s$
yields $\kro\simeq 0.7$, in
agreement with the exact result \cite{schulz_conductivite_1d}. We then
study how inter-chain coherence develops as temperature is lowered for coupled chains with $\tp/W= 0.14$.
Fig.~\ref{kro-vs-T} displays the effective $\kro$ as a function of
temperature: a clear crossover from a LL (with $\kro\simeq 0.7$)
to a FL ($\kro=1$) is seen as temperature is lowered below
$T^*/W\simeq 1/44 $.
According to \cite{bourbonnais_rmn,bourbonnais_couplage}, the crossover scale
is given by 
$T^* = \frac{\tp}{\pi}C (\tp/t)^{\theta/(1-\theta)}$,
with $\theta=(\kro+1/\kro)/4-1/2$ ($\simeq 0.03$ here). The interactions
reduce its value compared to the naive one $\tp/\pi$.
We cannot meaningfully test this renormalization effect because of the small
values of $\theta$ in the Hubbard model. Our results are
consistent with $T^* \simeq C t_\perp/\pi$ with a 
proportionality constant  $C\simeq 0.5$.
In the low temperature FL regime, we find
that the location of the FS is essentially unaffected by
interactions. The quasi-particle residue $Z$ is not uniform along
the FS. Its dependence on $\kp^F$ is displayed in
Table~\ref{tab1doped}. $Z$ is larger for $\kp$ close
to $\pi$, while it is somewhat smaller for small $\kp$. This is
not a very big effect however, and it would be inappropriate to
speak of ``hot spots'' \cite{zheleznyak_hot_spots} in that region
(even though it may be indicative that such a phenomenon appears
for stronger interactions). Note that the single-chain
approximation $\Sigma=\Sigma_{1D}$ used by many authors
would lead to a much stronger variation of $Z$ along the FS.
Our results, in agreement with those
of \cite{arrigoni_tperp_resummation}, show that 
$Z$ is
more uniform than this approximation would suggest.
\begin{table}
\caption{QP weights $Z(\kp)$ for different points on the FS (doped
case: $n\simeq 0.8$, $U/W=1.0$, $\tp/W=0.14$). \label{tab1doped}}
\begin{ruledtabular}
\begin{tabular}{lllllll}
$\kp/\pi$ & 0.19 & 0.36 & 0.50 & 0.62 & 0.74 & 0.91 \\
\hline
$Z(\kp)$  & 0.62  & 0.65 & 0.69 & 0.72 & 0.75 & 0.78
\end{tabular}
\end{ruledtabular}
\end{table}

\begin{figure}
\centerline{\includegraphics[width=\figwidth]{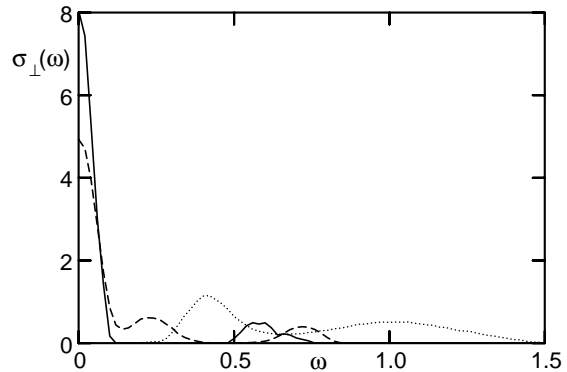}}
\caption{\label{fig-conduc} Inter-chain optical conductivity at
half-filling for $U=0.65 W$, $\beta=40/W$ and
$\tp=0.14$ and $0.07$ (solid and dashed lines) and
for $U=1.0W$, $\beta=40/W$ and $\tp=0.$ (dotted line).}
\end{figure}

\begin{figure}
\centerline{\includegraphics[width=\figwidth]{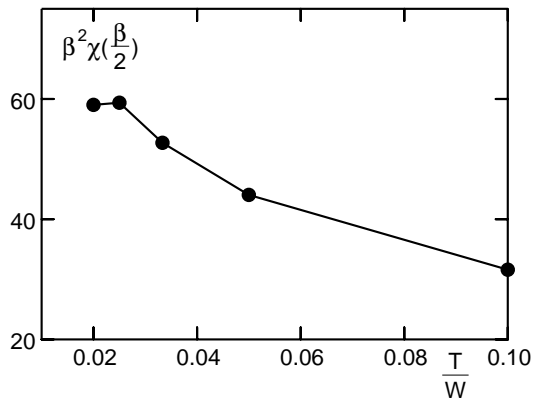}}
\caption{\label{T1T_mu0} $1/(T_1 T)$ versus T in the
half-filled case with $\tp/W=0.11$, $U/W=0.65$. The crossover
from a FL at low-T to a LL for $T/W\geq 0.025$ is
clearly apparent.}
\end{figure}

\begin{figure}
\centerline{\includegraphics[width=\figwidth]{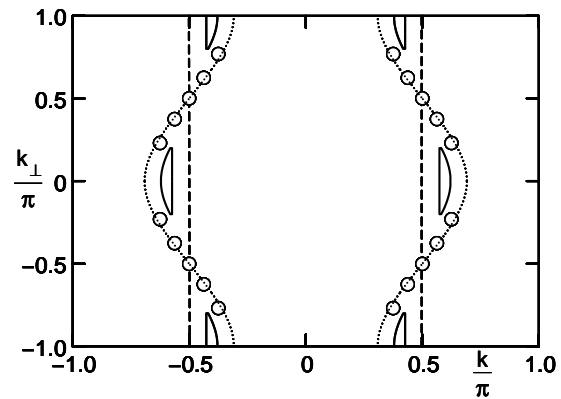}}
\caption{\label{FS-plot} FS in the half-filled case
with $\tp/W=0.14$, $U/W=0.65$ (circles), compared
to the FS of the non-interacting case (dotted line) and of the
purely 1d case ($\tp=0$ -dashed-.). The solid line depicts
schematically the Fermi surface obtained when making the
approximation $\Sigma=\Sigma_{1D}$ \cite{essler_mott_rpa}.}
\end{figure}

We now turn to the commensurate (half-filled) case, and present
results for $U=0.65\,W$. By fitting the self-energy to an
analytical form, we first checked that our numerical results are
consistent with a Mott insulating state with a gap $\Delta_{1D}/W=
0.1$, in agreement with exact results. Turning on $\tp$, we
estimate the $\tp$-dependence of the gap by performing simulations
at one of the lowest accessible temperatures $T/W=1/40$.
The gap vanishes for $\tp/W\simeq 0.07$. Hence our method captures the
insulator-to-metal transition induced by
transverse hopping.
\begin{table}
\caption{Effective $\kro$ at half-filling, as a
function of $\tp/W$ for $U/W=0.65$ and $T/W=1/40$.
\label{table-half}}
\begin{ruledtabular}
\begin{tabular}{llllllll}
$\tp/W$ & 0.00 & 0.04 & 0.07 & 0.11 & 0.14 & 0.16 & 0.18 \\
\hline
$\kro$  & 0.00  & 0.02 & 1.01 & 1.09 & 1.07 & 1.06 & 1.04
\end{tabular}
\end{ruledtabular}
\end{table}
The effective $\kro$, shown in Table~\ref{table-half} is
also a clear indicator of the MI
($K_\rho=0$) to FL ($K_\rho=1$) transition, in agreement with
the qualitative expectations of Fig.~\ref{Phase_diagram}. At this
low temperature, the intermediate LL regime is too narrow to be
visible. The location of the transition is in
reasonable agreement with the naive criterion $\Delta_{1D} \sim
t_\perp^*$
\cite{giamarchi_mott_shortrev,vescoli_confinement_science,
tsuchiizu_confinement_spinful_refs}.
Fig.~\ref{fig-conduc} displays the inter-chain
optical conductivity for several values of $\tp$. In the MI phase,
$\sp(\omega)$ shows a gap, followed by 
an onset of absorption starting at approximately 
the gap and
extending up to a scale of order $U$, where a broad second peak is
apparent.
A low-frequency Drude peak develops as
the insulator to metal transition is crossed. Close
to the transition, the weight in the Drude peak is small, while
the Hubbard band feature is still clearly visible and carries a
significant part of the spectral weight.
The results for the NMR relaxation rate $1/T_1$ are displayed in
Fig.~\ref{T1T_mu0} for a value of $\tp/W=0.11$.
At low temperatures the Korringa law $1/(T_1 T)= const$ is
recovered, while at higher temperatures $1/T_1T$ is temperature
dependent. The observed T-dependence is consistent with $1/(T_1
T) \simeq T^{\kro-1}$ \cite{bourbonnais_rmn} with a T-dependent exponent $\kro \le 1$,
indicative of a LL evolving gradually into a FL (cf. Fig.~\ref{Phase_diagram}).
As for the doped case, the FS is nearly
indistinguishable from the non-interacting case (see Fig.~\ref{FS-plot}). The QP
residue $Z_{\kp}$ depends only very weakly on the location on the
Fermi surface as shown in Table~\ref{tab1half}.
\begin{table}
\caption{QP weights $Z(\kp)$ for different points on the
FS (half-filled case, $\tp=0.14 W$, $U/W=0.65$).
\label{tab1half}}
\begin{ruledtabular}
\begin{tabular}{llllll}
$\kp/\pi$ & 0.23 & 0.38 & 0.50 & 0.62 & 0.77 \\
\hline
$Z(\kp)$  & 0.79  & 0.77 & 0.76 & 0.77 & 0.79
\end{tabular}
\end{ruledtabular}
\end{table}
The data may suggest very shallow minima, which in contrast to the
doped case, are found at $\kp\sim \pm \pi/2$, corresponding
to a vanishing inter-chain kinetic energy.
We would like to contrast our findings with the conclusions drawn
from the ``single-chain'' (RPA) approximation
\cite{wen_coupled_chains,essler_mott_rpa}: $\Sigma=\Sigma_{1D}$. As recently pointed out
\cite{essler_mott_rpa}, this approximation predicts that the FS
close to the metal-insulator transition should consist of
disconnected electron and hole ``pockets'', as depicted
schematically on Fig.~\ref{FS-plot}. This is because the
self-energy of a MI diverges at low-frequency for $k=\pi/2$, so
that no FS point can correspond to this value of $k$. In contrast,
our data find a conventional connected FS down to the transition
point (up to our numerical accuracy). This
shows that it is crucial, as done in \cDMFT, to take into
account the feedback effects of the interchain hopping in the
self-energy.

In conclusion, we have shown in this paper that the \cDMFT
approach is a tool of choice for the study of the crossovers and
insulator to metal transitions induced by transverse hopping in
quasi one-dimensional systems. Several features found in our study
are qualitatively reminiscent of experimental observations on
organic conductors, such as: the deconfinement transition itself,
the crossover from a LL at high-T to a FL at low-T (as revealed
e.g in NMR) and most significantly the coexistence of a Drude
feature with small spectral weight and of a Hubbard-band feature
in optical conductivity.
Extensions of the model including on-site and
nearest neighbor interactions would allow to study the
deconfinement transition at quarter-filling and thus
to make a more realistic \cite{giamarchi_mott_shortrev,schwartz_electrodynamics}
comparison with experimental results.
Though this is computationally much more demanding,
we hope to use \cDMFT in this
context in future work.
\begin{acknowledgments}
Acknowledgements: We acknowledge useful discussions with
A.~Tsvelik, F.~Essler, D.~Jerome and C.~Bourbonnais.
This research has been supported by a Marie Curie Fellowship of the
European Community Programme ``Improving Human Potential'' under
contract number MCFI-2000804 and a grant of supercomputing time at
NIC J{\"u}lich.
\end{acknowledgments}

\end{document}